\documentclass{ws-p9-75x6-50}

\def\eps@scaling{.95}%
\newcommand\epsscale[1]{\gdef\eps@scaling{#1}}%

\newcommand\plottwo[2]{%
 \centering
 \leavevmode
 \columnwidth=.45\columnwidth
 \includegraphics[natwidth={\eps@scaling\columnwidth}]{#1}%
 \hfil
 \includegraphics[natwidth={\eps@scaling\columnwidth}]{#2}%
}%

\begin{document}

\title{Preliminary Results from the Cosmic Background Imager}

\author{B.S. Mason, J.K. Cartwright, S. Padin, T.J. Pearson,
A.C.S. Readhead, M. Shepherd, J. Sievers, P. Udomprasert}

\address{105-24 Caltech \\
Pasadena, CA 91125 \\
Email:  {\tt bsm@astro.caltech.edu}}

\maketitle

\abstracts{
The Cosmic Background Imager (CBI) is a 13-element interferometer designed
to image intrinsic anisotropies in the cosmic microwave background (CMB)
on arcminute scales.  A review of the capabilities of the instrument is
presented, together with a discussion of observations which have been
taken over the past 9 months from the Atacama desert of Chile.  We present
preliminary high-resolution mosaiced images of the CMB obtained from
recent CBI data and discuss topics which the CBI will address in the near
future.}

\section{Introduction and Overview of the CBI}

Anisotropies in the Cosmic Microwave Background (CMB) contains a
wealth of information about fundamental cosmological parameters
\cite{White_et_al}, as well as providing a direct link to theories of
high-energy physics \cite{KK}.  Inspired by the possibility of
accurate determinations of classical cosmological parameters,
experiments have sought and detecteted the most prominent large-scale
CMB anisotropies due to the first Doppler peak ({\it e.g.},
\cite{Miller_et_al_1999,Leitch_et_al_2000,boom}).  Smaller angular
scales encode information from causal processes which are independent
of the location of the Doppler peaks.  The primary mission of the CBI
is to measure the CMB anisotropy spectrum in this relatively
unexplored regime.

\begin{figure}[t]
\begin{center}
\epsfxsize=2in 
\epsfbox{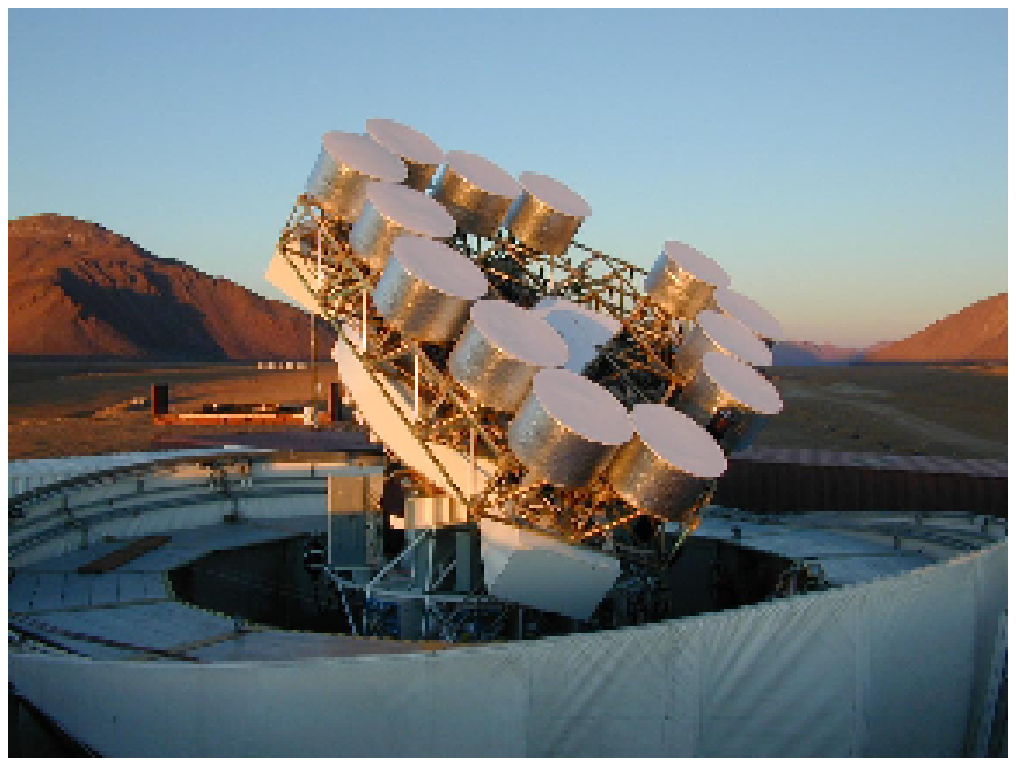}  
\hfill
\epsfxsize=2in
\epsfbox{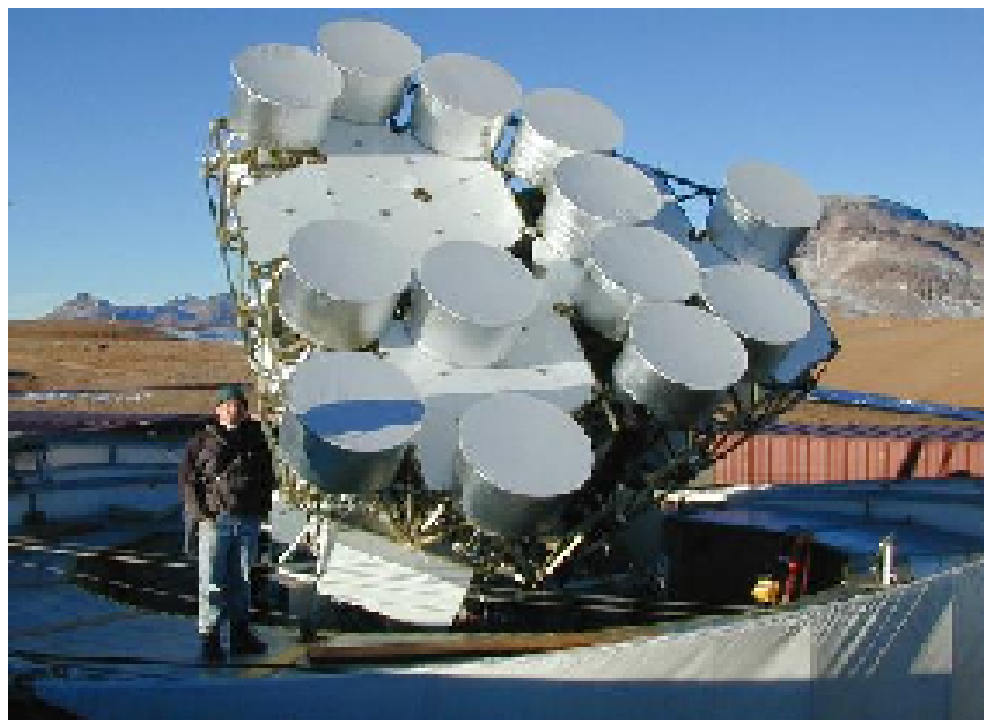}
\end{center}
\caption{The CBI in its test configuration (left) providing uniform
$uv-$coverage and easy access to all receivers, and in a configuration
providing more short baselines for mosaic and Sunyaev-Zel'dovich
Effect observations (right).  \label{fig:cbipic}}
\end{figure}

The CBI is an interferometric array of 13 0.9-meter diameter antennas
mounted on a 6-meter steerable platform, and operating in 10 1-GHz
bands between 26 and 36 GHz ($\sim 1 \, {\rm cm}$).  Configurations
available to the CBI yield synthesized beamwidths ranging from $4'$ to
$15'$ (FWHM).  Interferometry confers the significant advantage,
relative to total power or beam-switched single-dish methods, of
providing a {\it direct} measurement of $C_{\ell}$ on a scale
determined by the baseline length.  The range of baselines available
to the CBI correspond to $300 < \ell < 3000$.  By changing the array
configuration of the CBI, the instrument's sensitivity can be
optimized for varying ranges of $\ell$.  The primary beam width $44'$
(FWHM at 30 GHz) implies a resolution $\delta \ell \sim 420$ (FWHM);
this can be significantly improved by mosaicked observations.  Two
configurations of the CBI are shown in Figure~\ref{fig:cbipic}.

The CBI will be capable of providing the first single-experiment
measurement of the photon diffusion scale which is a fundamental
prediction of current theories of CMB anisotropy formation\cite{silk}.
Results at the low end of the range of $\ell$ available to the CBI
will provide an important check on other power spectrum determinations
in this range such as RING5M and BOOMERANG/MAXIMA, which are total
power experiments.  Two other interferometers are DASI\cite{halverson}
and the VSA\cite{grainge}, both of which operate at lower $\ell$ than
the CBI.  DASI is a sister experiment to the CBI with many design
elements in common, and these two interferometers together cover the
range $100 < \ell < 3000$.

\section{Observing Strategy and Preliminary Results}
Construction of the CBI was begun in August 1995 in Pasadena, CA on
the Caltech campus, and completed in January 1999.  After a period of
test observations in Pasadena, the telescope was disassembled and
shipped in August 1999 to its site high in the Chilean Andes.  This
site, at an altitude of 5000 meters, was chosen in order that our
sensitivity not be limited by atmospheric water vapor emissions.
First light in Chile was achieved November 1, 1999, and routine
observations have been taken from January 2000 to the present.

Our observing strategy is dictated by the fact that, on the shortest
CBI baselines, the ground produces significant correlated signals.  To
remove these, pairs of fields are observed over identical ranges in
azimuth and elevation.  The difference between these data cancels
ground-based signals.  This differencing also controls other possible
systematics due, {\it e.g.}, to correlator offsets or antenna
cross-talk.  From January through April 2000, observations of two
pairs of such fields were taken in a test configuration providing a
maximally uniform distribution of baseline lengths.  In all $\sim 160$
hours of integration were obtained, resulting in a robust detection of
CMB anisotropy on scales from $\ell = 300$ to $\ell = 1500$.  Results
from these observations have been accepted for publication \cite{padin}.

In order to remove discrete radio sources, the dominant foreground to
CMB measurements at our frequency and resolution, dedicated
observations of NVSS\cite{condon} selected sources are conducted at 30
GHz with the OVRO 40-meter telescope.  Sources detected by the
40-meter are subtracted directly from the CBI data.

\begin{figure}[t]
\begin{center}
\epsfxsize=2in 
\epsfbox{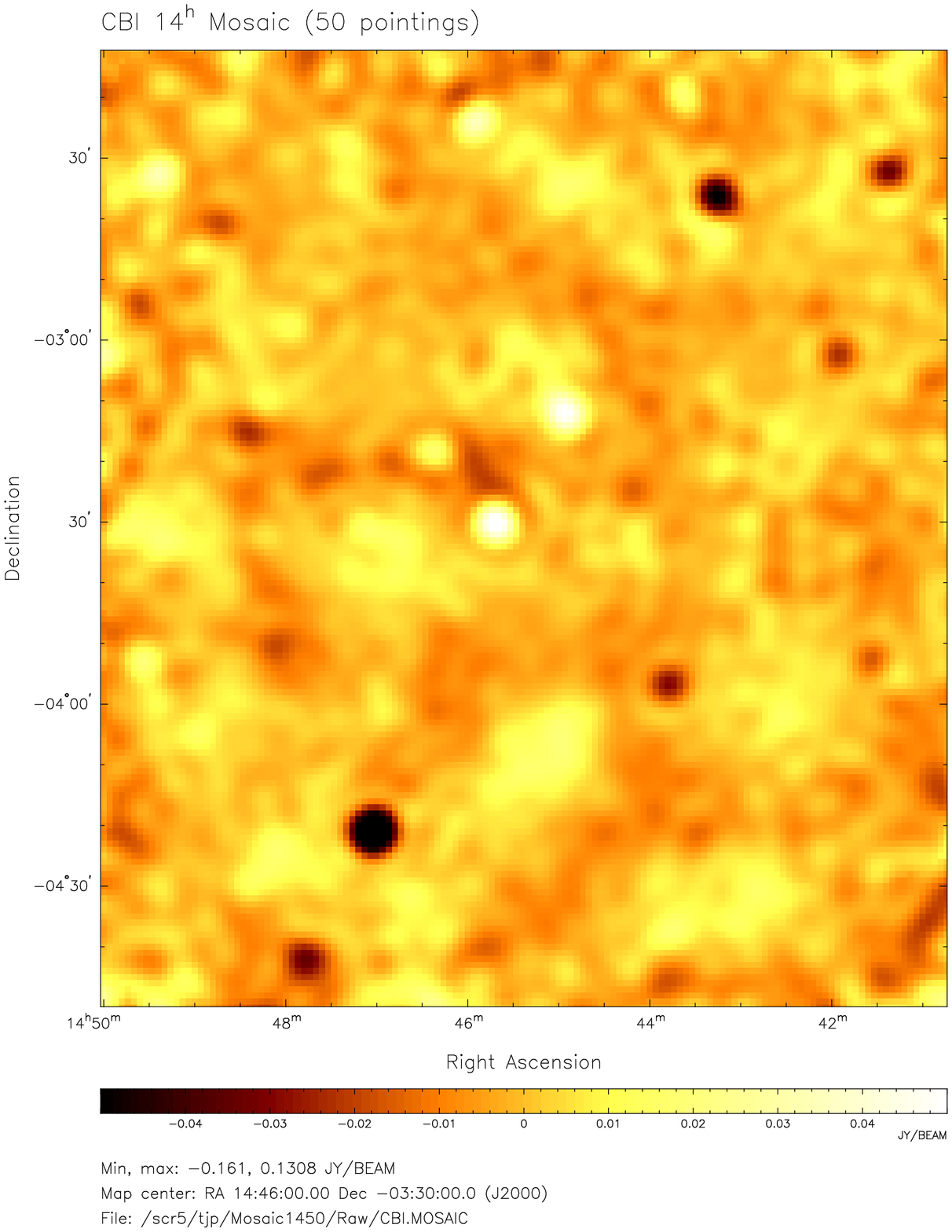}
\hfill
\epsfxsize=2in 
\epsfbox{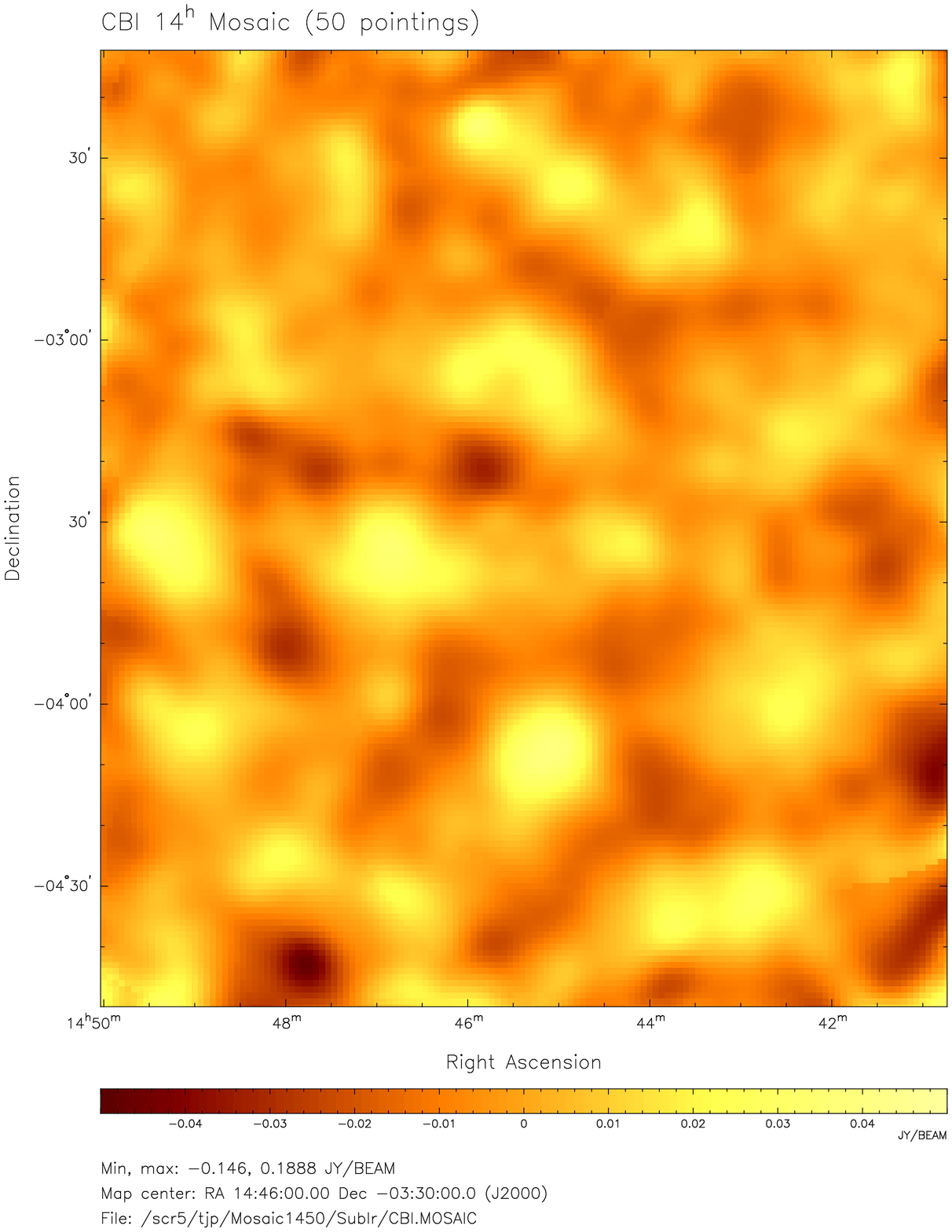}
\end{center}
\caption{Images from observations of one mosaic field.  
On the left, the raw mosaic data are shown, including contaminating
discrete sources.  The image on the right has had discrete sources
removed {\it via} independent observations with the OVRO 40-m, and the
long baseline data are downweighted to emphasize the large-scale
structure in the field.  Most of this is due to intrinsic CMB
fluctuations. \label{fig:mosaic}}
\end{figure}

Observations of three $5^{\circ} \times 3^{\circ}$ mosaic fields are
in progress.  When complete, these results will improve our resolution
in $\ell$ by a factor of $3 - 4$.  Figure~\ref{fig:mosaic} shows a
difference image of the CMB from preliminary observations of one of
our three mosaic fields before and after the removal of discrete
source contamination.

\section{Other Science with the CBI}

The CBI has one cross-polarized antenna.  This will enable stringent
limits to be placed on the CMB polarization in the vicinity of the
first Doppler peak in the polarization power spectrum.  Polarization
data have been acquired on the CBI fields from January 2000 through
the present.

The range of angular scales available to the CBI are also well-suited
to measuring the Sunyaev-Zeldovich Effect (SZE) in nearby galaxy
clusters.  A campaign to determine ${\rm H_0}$ from observations of
the SZE in a sample of 20 $z<0.1$ clusters is underway with the CBI.
This campaign has the feature, unique among current SZE programs, of
selecting targets from an orientation-unbiased sample
\cite{Udomprasert,Mason_and_Myers}.  The large sample size is
important for reducing the effects of intrinsic CMB anisotropy, and
for understanding possible X-ray modelling systematics associated with
clusters at a range of dynamical states.

\section*{Acknowledgments}
The CBI is a project of the California Institute of Technology, in
collaboration with the Universidad de Chile.  The construction of the
CBI has been made possible by the generous support of the California
Institute of Technology, Ronald and Maxine Linde, Cecil and Sally
Drinkward, and grants from the National Science Foundation (awards
AST-9413935 and AST-9802989).


\begin{thebibliography}{99}


\bibitem{White_et_al}White, M., Scott, D. \& Silk, J., \Journal{Ann. Rev. A \& A}{32}{319}{1994}

\bibitem{KK}Kamionkowski, M. \& Kosowski, A., \Journal{Ann. Rev. Nucl. Part. Sci.}{49}{77}{1999}

\bibitem{Miller_et_al_1999}Miller, A.D. {\it et al.}, \Journal{ApJ}{524}{L1}{1999}

\bibitem{Leitch_et_al_2000}Leitch, E.M. {\it et al.}, \Journal{ApJ}{532}{37}{2000}

\bibitem{boom}de Bernardis, P. {\it et al.}, \Journal{Nature}{404}{955}{2000}

\bibitem{silk}Silk, J., \Journal{ApJ}{151}{459}{1968}

\bibitem{halverson}Halverson, N.W., Carlstrom, J.E., Dragovan, M., Holzapfel, W.L., and Kovac, J., \Journal{SPIE}{3357}{416}{1998}

\bibitem{grainge}Grainge, K. {\it et al.}, this volume.

\bibitem{padin}Padin, S. {\it et al.},  ApJL {\it in press} ({\tt astro-ph/0012212})

\bibitem{condon}Condon, J. {\it et al.}, \Journal{AJ}{116}{1693}{1998}

\bibitem{Udomprasert}Udomprasert, P.S., Mason, B.S., \& Readhead,
A.C.S., to appear in {\it Constructing the Universe with Clusters of
Galaxies}, ed. F. Duret and D. Gerbel, ``The Sunyaev-Zel'dovich Effect
with the Cosmic Background Imager.''

\bibitem{Mason_and_Myers}Mason, B.S. \& Myers, S.T., \Journal{ApJ}{540}{614}{2000}

\end{thebibliography}
\end{document}